\documentclass{article}
\usepackage{spconf,amsmath,graphicx}
\usepackage{bm, amssymb}
\usepackage{multirow}
\usepackage{pgfplots}
\usepackage{import}
\pgfplotsset{compat=1.18}
\usepackage{flushend} 
\usepackage{acronym}
\acrodef{ASR}  {automatic speech recognition}
\acrodef{BLSTM}{bidirectional long short-term memory}
\acrodef{CEC}  {Clarity Enhancement Challenge}
\acrodef{CNN}  {convolutional neural network}
\acrodef{CPC}  {Clarity Prediction Challenge}
\acrodef{CPC1} {Clarity Prediction Challenge 1}
\acrodef{CPC2}{Clarity Prediction Challenge 2}
\acrodef{dBHL} {decibels hearing level}
\acrodef{DFT}  {discrete Fourier transform}
\acrodef{GAN}  {Generative Adversarial Network}
\acrodef{h2h}  {human-to-human}
\acrodef{h2m}  {human-to-machine}
\acrodef{HA}   {hearing aid}
\acrodef{HL}   {hearing loss}
\acrodef{HLS}  {hearing loss simulator}
\acrodef{HASPI}{Hearing-Aid Speech Perception Index}
\acrodef{HSR}  {human speech recognition}
\acrodef{MBSTOI}{Modified Binaural Short-Time Objective Intelligibility}
\acrodef{MOS}  {mean opinion score}
\acrodef{MMSE} {minimum mean squared error}
\acrodef{MSE}  {mean squared error}
\acrodef{NN}   {neural network}
\acrodef{OLA}  {Overlap-Add}
\acrodef{PESQ} {Perceptual Evaluation of Speech Quality}
\acrodef{RMSE} {Root Mean Square Error}
\acrodef{SE}   {speech enhancement}
\acrodef{SI}   {speech intelligibility}
\acrodef{SNR}  {Signal to Noise Ratio}
\acrodef{SPIN} {Speech In Noise}
\acrodef{SSSR} {self-supervised speech representation}
\acrodef{STFT} {short time Fourier transform}
\acrodef{STOI} {Short-Time Objective Intelligibility}
\acrodef{VAD}  {Voice Activity Detector}
\acrodef{WER}  {Word Error Rate}
\acrodef{XLSR} {Cross-Lingual Speech Representation}

\usepackage[citecolor=black,
            colorlinks=true,
            linkcolor=black,
            urlcolor  = black,
            pdftitle={Non-Intrusive Speech Intelligibility Prediction for Hearing-Impaired Users using Intermediate ASR Features and Human Memory Models},
            pdfauthor={Rhiannon Mogridge, George Close, Robert Sutherland, Thomas Hain, Jon Barker,
            Stefan Goetze, Anton Ragni},
            pdfkeywords={speech recognition, human-computer interaction, computational paralinguistics},
            pdfsubject={speech recognition, human-computer interaction, computational paralinguistics}]{hyperref}
\usepackage[english]{babel}
\addto\extrasenglish{}
\addto\extrasenglish{}
\addto\extrasenglish{}
\addto\extrasenglish{}
\addto\extrasenglish{}
\addto\extrasenglish{}

\title{Non-Intrusive Speech Intelligibility Prediction for Hearing-Impaired Users using Intermediate ASR Features and Human Memory Models}
\name{Rhiannon Mogridge, George Close, Robert Sutherland, Thomas Hain, Jon Barker, Stefan Goetze, Anton Ragni\thanks{This work was supported by the Centre for Doctoral Training in Speech and Language Technologies (SLT) and their Applications funded by UK Research and Innovation [grant number EP/S023062/1]. This work was also supported by Toshiba and WS Audiology.}}
\address{Speech and Hearing Group, Dept.~of Computer Science, University of Sheffield, UK}
\begin{document}
\ninept
\maketitle
\begin{abstract}
Neural networks have been successfully used for non-intrusive speech intelligibility prediction. Recently, the use of feature representations sourced from intermediate layers of pre-trained self-supervised and weakly-supervised models has been found to be particularly useful for this task. This work combines the use of Whisper ASR decoder layer representations as neural network input features with an exemplar-based, psychologically motivated model of human memory to predict human intelligibility ratings for hearing-aid users. Substantial performance improvement over an established intrusive HASPI baseline system is found, including on enhancement systems and listeners unseen in the training data, with a root mean squared error of $25.3$ compared with the baseline of $28.7$. 
\end{abstract}

\section{Introduction}

Hearing loss is a widespread problem that affects approximately $466$ million people worldwide (around $6\%$ of the world population), though this problem is only expected to grow; by $2030$ it is predicted to impact $630$ million people worldwide \cite{world2018addressing}. Age correlates with the chance that a person will be affected by hearing impairment \cite{RGA11}, and the population is expected to age. From  $2015$ to $2050$, the proportion of the population aged over $60$ is expected to almost double, rising from $12$\% to $22$\% \cite{aging_stats}. As hearing impairment typically worsens in an individual, their ability to make intelligible  the speech that they hear decreases. 

Successful development of \ac{HA} technology \cite{Doclo_HA_Overview,GXRRA10} requires assessment, which is expensive and time-consuming when conducted by human listeners \cite{falk2015objective,WKJ+14}. Automated intelligibility metrics, which are designed to mimic human assessment, are more cost-effective and can also be used as training objectives~\cite{Voelker_SI_2015,CSS+19}. Developing and improving estimators for such metrics is therefore essential.

The \ac{CPC2} \cite{barker24_icassp} builds on the prior \ac{CPC1} \cite{barker22_interspeech}. It provides a comparatively large dataset of noisy audio processed by hearing aid systems, each with an associated intelligibility score obtained from listening tests with hearing-impaired human listeners.

The challenge task is to predict the intelligibility score of hearing-impaired listeners given the audio and some additional information such as a representation of the hearing loss of the listener. The challenge has both a non-intrusive track, where only the \emph{noisy} signal processed by the hearing aid can be used, and an intrusive track, where a \emph{clean} version of the input audio can also be used.

In this work, an intelligibility estimator for the non-intrusive challenge task is proposed, which makes use of a feature transformation of the audio signal, derived from intermediate representations of a pre-trained \ac{ASR} model (cf.~\autoref{sec:features}). This is combined with a model of human memory, which has its origins in the field of human psychology, to predict the intelligibility score of hearing aid output audio (cf.~\autoref{ssec:Exemplar-informedModel}). 

The remainder of this work is organised as follows: \autoref{sec:challenge} briefly describes the \ac{CPC2}, including the dataset and baseline model. \autoref{sec:architecture} covers the proposed features and model architecture. Results are presented in \autoref{sec:Results} and \autoref{sec:Conclusions} concludes the paper.

\section{Clarity Prediction Challenge 2}
\label{sec:challenge}

The Clarity Project runs two challenges in sequence with the aim of improving \ac{HA} technology: the \ac{CEC} and the \ac{CPC}. The \ac{CEC} objective is to design systems to enhance noisy signals for hearing-impaired listeners. The \ac{CPC} objective is to assess the intelligibility of the \ac{CEC} systems. 

\subsection{Challenge data}

The \ac{CPC2} data consists of tuples of a speech signal $\hat{s}[n]$  and its corresponding correctness value $i$, obtained from listening tests with hearing-impaired listeners. 


The signal $\hat{s}[n]$ is the enhanced outputs of hearing aid systems with binaural input $x[n]$, being  an artificially corrupted version of clean reference audio $s[n]$ with additive noise $v[n]$. The correctness value $i$ is the percentage of words which a hearing-impaired listener was able to correctly reproduce from the speech signal $\hat{s}[n]$ they listened to. The challenge data also contains additional information such as left/right ear's representations of the listeners' hearing loss as audiograms $\mathbf{a}_l$ and $\mathbf{a}_r$. All audio signals are stereo with a left and right channel. 

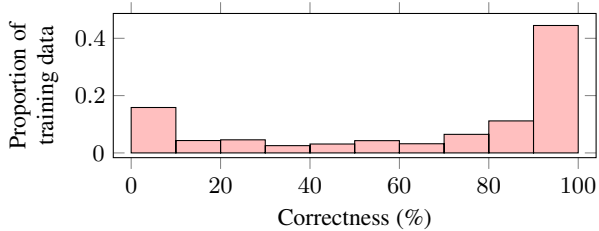
\begin{figure}[t]
    \centering
    \begin{tikzpicture}
        \begin{axis}[
            width=8cm,
            height=3.5cm,
            ybar,
            enlargelimits=0.1,
            bar width=10,
            ylabel style={align=center, text height=8mm},
            yticklabel style={text width=4mm, align=right},
            ylabel={Proportion of\\training data},
            xlabel={Correctness (\%)},
            ]
            \addplot [fill=pink] coordinates {  
                (5,0.1585) (15,0.0432) (25,0.0457) (35,0.0254) (45,0.0311) (55,0.0428) (65,0.0320) (75,0.0648) (85,0.1117) (95,0.4447) 
            };
        \end{axis}
    \end{tikzpicture}
    \caption{Distribution of true correctness values in the training data.}
    \label{fig:data_dist}
\end{figure}

The resulting data is partitioned into three train sets, each paired with a disjoint evaluation set. Each evaluation set covers listeners and hearing aid enhancement systems which are unseen in its corresponding training set, meaning that prediction models need to generalise to unseen listeners and systems. Set $1$ has a training set of size $8599$ and an evaluation set of size $305$ audio samples. Set $2$ has a training set of size $8135$ and an evaluation set of size $294$.  Set $3$ has a training set of size $7896$ and an evaluation set of size $298$. There are around 40 hours of audio in total. Audio with very low intelligibility (correctness 0) and very high intelligibility (correctness 100) are over-represented, as shown in \autoref{fig:data_dist}.


\subsection{Prior approaches}
A number of different approaches were taken in the first iteration of the challenge in CPC1~\cite{barker22_interspeech}. The best performing non-intrusive approach \cite{tu22b_interspeech} uses an uncertainty measure derived from state-of-the-art \ac{ASR} systems as a proxy for human intelligibility, finding a strong correlation between the two measures. Other successful approaches ~\cite{zezario2022mbinet,close2023non} make use of powerful feature representations derived from \acp{SSSR} as inputs to neural speech intelligibility prediction models, while others use neural network structures which have been shown to be useful in the related task of human speech quality rating prediction~\cite{close22_interspeech}.

\ac{CPC2} differs from \ac{CPC1} in that its evaluation sets are disjoint in terms of listener and hearing aid system relative to its training sets. This means that some of the better-performing approaches to \ac{CPC1}, which operated effectively as predictors of the hearing aid system, were not at all useful in \ac{CPC2}, which was discovered in early experiments for this work with the \ac{CPC2} data. As such, our proposed system for \ac{CPC2} builds on the best-performing approaches to \ac{CPC1}, while ensuring that it can generalise to unseen data. 
\subsection{Challenge baseline}

The baseline system provided by the challenge organisers \cite{barker24_icassp} makes use of the \ac{HASPI}, version 2 \cite{kates2021hearing}. This is an intrusive system that makes use of both the enhanced noisy signal $\hat s[n]$ and the clean speech signal $s[n]$. A \ac{HASPI} score is computed for both the left and right ear signals and logistic regression is used to predict the correctness scores from the higher \ac{HASPI} score.

\section{System architecture}
\label{sec:architecture}
This section describes the proposed approach to the \ac{CPC2} task. The approach consists of neural networks which take a recent \ac{ASR} derived representation of the hearing aid output signal $\hat{s}[n]$ as input and return a prediction $\hat i$ of the correctness value.

\subsection{Features}
\label{sec:features}

The Whisper model \cite{whisper} is an \ac{ASR} model pre-trained on $680,000$ hours of multi-lingual data for tasks including English and non-English speech transcription and voice activity detection. The model architecture is an encoder-decoder transformer \cite{vaswani2017attention}, with $12$ encoder layers and $12$ decoder layers for the \emph{small} Whisper model\footnote{\url{https://huggingface.co/openai/whisper-small}} used in this work. 

Given the enhanced speech signal $\hat{s}[n]$, with corresponding spectrogram representation $\mathbf{\hat{S}}$, the input to the proposed intelligibility prediction model is the output of the $12$ decoder layers of the Whisper model. The Whisper decoder layers output represents word-level features, with dimension $W \times 768 \times 12$, where $W$ is the predicted number of words in the utterance, $768$ is the feature dimension of each decoder layer for the \emph{small} Whisper model, and $12$ is the number of decoder layers. Note that while $W$ varies from utterance to utterance, for a given utterance it will remain fixed through the decoder layers. The parameters of the Whisper model are frozen during training of the metric prediction model described below, i.e., the Whisper model is used as a feature transform.


\subsection{Model structure}

\begin{figure}[t]
  \centering
  \includegraphics[width=0.98\columnwidth]{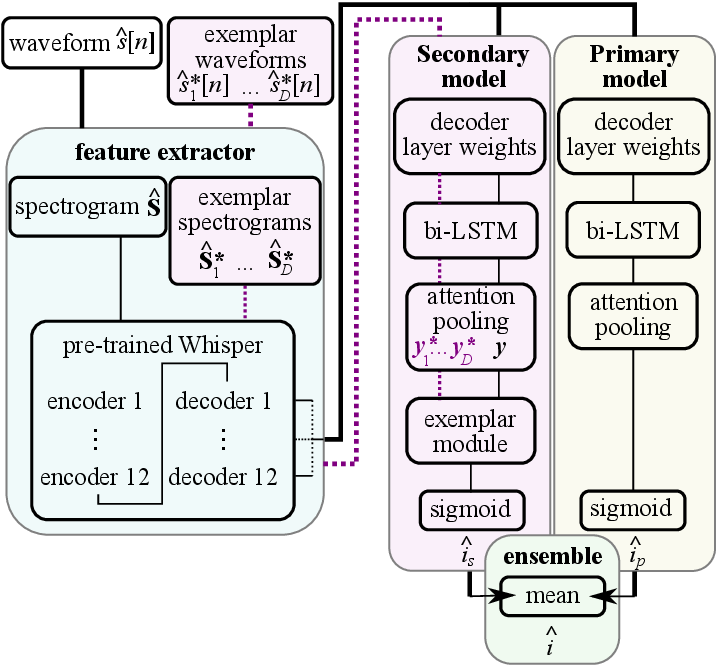}
  \caption{Model architecture of proposed \ac{SI} prediction.}
  \label{fig:architecture}
\end{figure}

An ensemble of two models for \ac{SI} prediction is used, as shown in \autoref{fig:architecture}, and the results of these two models are combined by averaging. 

A model structure following work on the \ac{CPC1} in \cite{close2023non} is chosen for the \emph{primary} \ac{SI} prediction network (cf.~\autoref{ssec:PrimaryModel}), depicted to the right in \autoref{fig:architecture}. This model structure has previously been successfully applied to the task of human quality label prediction \cite{nisqa_pretrained_ss} using pre-trained \ac{SSSR} representations as its input feature. This approach was also found to be useful for the \ac{CPC1} task~\cite{close2023non}, however generalization to unseen hearing aid systems was poor.  

The \emph{secondary} \ac{SI} prediction network incorporates an exemplar-informed module based on a simplified theory of human memory \cite{hintzman1984minerva2} (cf.~\autoref{ssec:Exemplar-informedModel}), and is shown in the middle-right of \autoref{fig:architecture}. Humans are believed to make use of specific examples, or exemplars, for memory-based tasks \cite{psyEx_rulePlusEx,humanRuleExemp:1998,humanExProt:2006,psyPersonaSwitch}. Humans are also able to non-intrusively assess speech signals, i.e.,~without direct reference. Since the challenge objective is to predict human responses, incorporating an exemplar-informed component may provide benefits or insight.

The output of the ensemble $\hat{i}$ for a given input signal $\hat{s}[n]$ is the mean of the outputs of the primary $\hat{i}_p$ and secondary systems $\hat{i}_s$.

\subsubsection{Primary SI prediction model}
\label{ssec:PrimaryModel}

The model structure uses a learnable weighted sum of the Whisper representations, implemented as a learnable linear layer with $12$ parameters, all initialised to $1$, followed by a softmax to ensure that the layer weights sum to $1$. This representation of dimension $W \times 768$ is then processed by $2$ \ac{BLSTM} layers with an input size of $768$ and a hidden layer size of $384$. Finally, an attention pooling feed-forward layer with sigmoid activation outputs to a single neuron which represents the primary predicted correctness value $i_p$ normalized between $0$ and $1$. The primary model has approximately $8.3$ M parameters.

\subsubsection{Secondary SI prediction model: exemplar-informed}
\label{ssec:Exemplar-informedModel}
The secondary model differs from the primary model in that the attention-pooling output feeds into an exemplar-informed module based on a simplified theory of human memory \cite{hintzman1984minerva2}. The exemplar-informed module incorporates a memory set of $D$ exemplars, which are speech signals $\hat{s}^*_1[n], ..., \hat{s}^*_D[n]$ drawn from the training data, with corresponding correctness values $i^*_1, ..., i^*_D$. The exemplars can be changed during training and for inference. Let $\bm y$ be the output of the attention pooling for input $\hat{s}[n]$ (see \autoref{fig:architecture}). The exemplars, $\hat{s}^*_1[n], ..., \hat{s}^*_D[n]$, are processed in the same way as the input, producing exemplar outputs from the attention pooling $\bm y^*_1, ..., \bm y^*_D$.  The output, $r$, of the exemplar module is given by
\begin{align}
    a &= \sum_{d=1}^D \frac{\bm f(\bm y) \cdot \bm g(\bm y^*_d)}{|| \bm f(\bm y) || \: || \bm g(\bm y^*_d) ||} i^*_d \\
    r &= h(a)
\end{align}
The functions $\bm f: \mathbb R^{768} \rightarrow \mathbb R ^{768}$, $\bm g: \mathbb R^{768} \rightarrow \mathbb R^{768}$ and $h: \mathbb R \rightarrow \mathbb R$ are all learned affine transformations. The value $a$ is a combination of the exemplar labels, weighted by their similarity to the input. This passes through a single linear neuron to produce $r$, and then through a sigmoid activation, which yields the secondary model's prediction, $\hat i_s$, normalised to fall between $0$ and $1$. The exemplar model has approximately $10$~M parameters.

\subsection{Experimental setup}

All audio used as input to Whisper is re-sampled to $16$kHz and padded/truncated to be 30 seconds long. In the case of the \ac{CPC2} data, all recordings were shorter than 30 seconds, so were padded, and downsampled from $32$kHz to $16$kHz. From this time-domain signal, the $80$ channel log magnitude Mel spectrogram is computed, using a window of $25$ms and a stride of $10$ms. During training, both left and right ear signals are used  as independent samples with the same correctness label. During inference, the signal that produces the highest correctness value is used to account for the better-ear effect \cite{better_ear}.

For each of the three splits, two listeners and two systems were randomly selected to form a disjoint validation set. All data with these listeners and systems were removed from the training set. A randomly selected non-disjoint validation set consisting of 10\% of the remaining training data was also formed. The majority of model selection and hyperparameter tuning was performed using these validation sets, to test generalisation to unseen listeners and systems. For the final models, the disjoint validation set and all listeners/systems associated with it were merged back into the training data to make the best use of resources. 

The primary and secondary models are trained separately with mean squared error loss. The primary model is trained for $25$ epochs with batch size $8$, learning rate $10^{-5}$ and weight decay $10^{-4}$. The secondary model is trained for $50$ epochs with learning rate $2 \times 10^{-6}$ and weight decay $10^{-4}$. During training and validation, $D = 8$ exemplars are chosen randomly from the training data for each minibatch.
\section{Results and discussion}
\label{sec:Results}

\autoref{tab:results} shows the results for the primary, secondary and ensemble models on each of the data splits, as well as the overall result for the \ac{CPC2} baseline. The validation sets contain enhancement systems and listeners seen in the corresponding training data. The evaluation sets are disjoint, containing only unseen enhancement systems and listeners. The primary, secondary and ensemble prediction networks all beat the baseline on all evaluation sets. 
\begin{table}[th]
  \caption{Validation and evaluation set results.}
  \label{tab:results}
  \centering
  \resizebox{\columnwidth}{!}{%
  \begin{tabular}{ l|cccc|cccc }
    \hline
    \multirow{3}{*}{\textbf{Model}} & \multicolumn{8}{c}{\textbf{RMSE}} \\
    & \multicolumn{4}{c}{\textbf{validation split}} & \multicolumn{4}{c}{\textbf{evaluation split}} \\
    & 1 & 2 & 3 & all & 1 & 2 & 3 & all \\
    \hline
    CPC2 baseline & & & & & & & & \textbf{28.7} \\
    Primary & 21.6 & 23.5 & 22.8 & \textbf{22.7} & 28.2 & 23.8 &  23.3 & \textbf{25.3} \\ 
    Secondary & 21.7 & 23.5 & 22.7 & \textbf{22.7} & 29.1 & 24.5 & 23.4 & \textbf{25.8} \\ 
    Ensemble & 21.6 & 23.4 & 22.7 & \textbf{22.5} & 28.6 & 23.9 & 23.2 & \textbf{25.3}\\ 
    \hline
  \end{tabular}
  }
\end{table}
\subsection{Primary and secondary models}
The primary and secondary systems show similar performance on the validation sets, with the ensemble of the two outperforming either. This is not replicated on the evaluation set, where the primary model outperforms the secondary model, and performs equivalently to the ensemble. 
\subsection{Performance by intelligibility}
\begin{figure}[ht]
    \centering
    \begin{tikzpicture}
        \begin{axis}[
            width=8cm,
            height=3.5cm,
            ymin=0,
            ybar,
            enlargelimits=0.1,
            bar width=10,
            ylabel style={align=center, text height=8mm},
            yticklabel style={text width=4mm, align=right},
            ylabel={{ }\\RMSE},
            xlabel={Correctness (\%)},
            ]
            \addplot [fill=pink] coordinates {  
                (5,16.98) (15,29.55) (25,28.93) (35,31.88) (45,32.20) (55,31.31) (65,35.15) (75,28.92) (85,22.07) (95,17.28) 
            };
        \end{axis}
    \end{tikzpicture}
    \caption{Model performance by true correctness.}
    \label{fig:correctness_bar}
\end{figure}
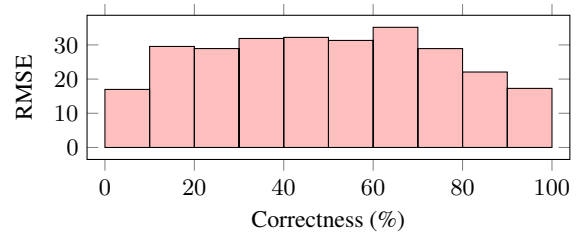
Figure~\ref{fig:correctness_bar} shows the performance of the ensemble model for different correctness values. The model performs well for very low intelligibility (0 correctness) and for very high intelligibility (100 correctness) but performs less well between the two extremes. This corresponds with the distribution of true correctness scores in the training data (see \autoref{fig:data_dist}), in which 0 and 100 correctness are over-represented. 

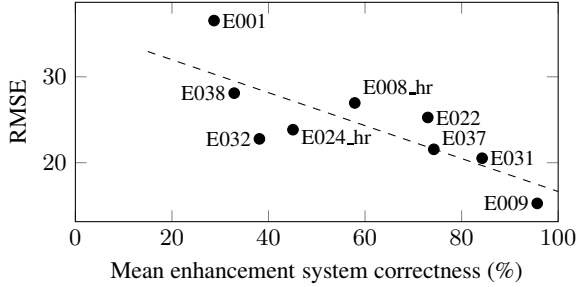
\begin{figure}[ht]
    \centering
    \begin{tikzpicture}
        \begin{axis}[
            width=8cm,
            height=4.5cm,
            xmin=0,
            xmax=100,
            ylabel={RMSE},
            xlabel={Mean enhancement system correctness (\%)}
            ]
            \addplot [black, dashed, domain=15:100, mark=none] {35.819-0.1917*\x};
            \addplot[
                scatter,
                black,
                mark=*,
                only marks,
                scatter src=explicit symbolic,
                visualization depends on=\thisrow{alignment} \as \alignment,
                nodes near coords,
                point meta=explicit symbolic, 
                every node near coord/.style={anchor=\alignment, font=\footnotesize}
            ]
            table[meta index=2] {
                x y label alignment
                28.722 36.5399182077136 E001 180
                57.872 26.95570662357407 E008\_hr 200
                95.659 15.267219660029815 E009 0
                72.994 25.265223948179752 E022 180
                45.073 23.842532789596014 E024\_hr 170
                84.250 20.522091276992835 E031 180
                38.123 22.774675285482896 E032 0
                74.237 21.560360435983355 E037 200
                32.900 28.096722666528382 E038 0
            };
        \end{axis}
    \end{tikzpicture}
    \caption{Model performance by mean hearing aid system correctness.}
    \label{fig:rmse_by_systemRMSE}
\end{figure}

\subsection{Model performance on unseen enhancement systems}
All the models show lower performance on Evaluation Set $1$. This appears to stem from the presence of audio enhanced by enhancement system E001 (the baseline in Clarity Enhancement Challenge 2) in this evaluation set. Audio enhanced by this system has an average correctness value of $28.7\%$, which is significantly lower than the average of the other two enhancement systems in Evaluation Set $1$, E022 and E031, which have average correctness values of $73.0\%$ and $84.3\%$, respectively.

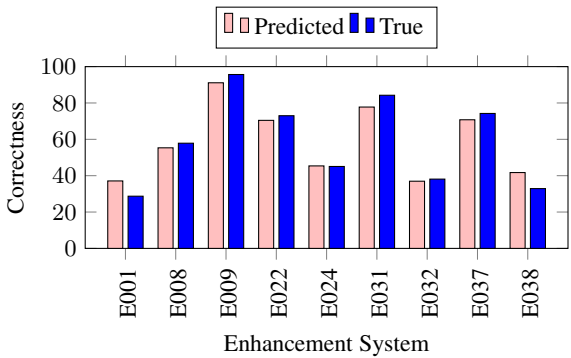
\begin{figure}[!ht]
    \centering
    \begin{tikzpicture}
        \begin{axis}[
            width=8cm,
            height=4cm,
            ymin=0,
            ymax=100,
            ybar,
            bar width=0.2cm,
            symbolic x coords={E001,E008,E009,E022,E024,E031,E032,E037,E038},
            xtick={E001,E008,E009,E022,E024,E031,E032,E037,E038},
            xticklabel style={text height=2ex,rotate=90},
            ylabel style={align=center, text height=8mm},
            ylabel={Correctness},
            xlabel={Enhancement System},
            legend columns = -1,
            legend style={at={(0.5,1.33)},anchor=north}
]
            \addplot[opacity=0.5,fill=pink] coordinates {
            (E001, 37.11)
            (E008, 55.33)
            (E009, 91.11)
            (E022, 70.44)
            (E024, 45.38)
            (E031, 77.76)
            (E032, 36.96)
            (E037, 70.76)
            (E038, 41.72)
            };
           \addplot[fill=blue, opacity=0.5] coordinates {
            (E001, 28.72)
            (E008, 57.87)
            (E009, 95.66)
            (E022, 72.99)
            (E024, 45.07)
            (E031, 84.25)
            (E032, 38.12)
            (E037, 74.24)
            (E038, 32.90)
            };
            \legend{Predicted, True};
        \end{axis}
    \end{tikzpicture}
    \caption{Performance of proposed prediction system depending on enhancement system}
    \label{fig:per-system}
\end{figure}

Although the proposed model can generalise to unseen enhancement systems, it predicts correctness less accurately on enhancement systems with lower performance. \autoref{fig:rmse_by_systemRMSE} shows the proposed model's performance by enhancement system correctness across all enhancement systems. There is a clear trend, with the proposed system more accurately predicting the correctness for better-performing enhancement systems which produce outputs with high correctness ratings. Conversely, the proposed system less accurately predicts outputs from poorly-performing enhancement systems which produce outputs with low correctness ratings. \autoref{fig:per-system} shows the predicted (left) and true (right) correctness across all enhancement systems, showing the proposed model overestimates the correctness of the two worst-performing enhancement systems, E001 and E038, while generally slightly underestimating the correctness of the better-performing enhancement systems. 

\subsection{Whisper layer weights}
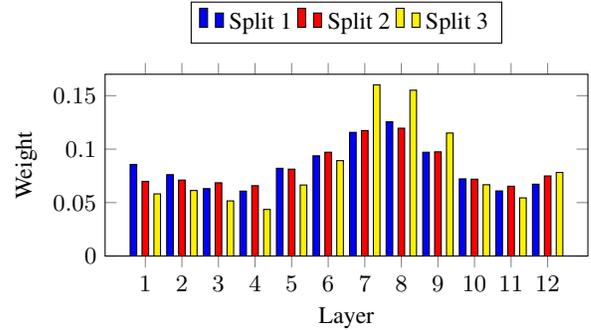
\begin{figure}[ht]
    \centering
    \begin{tikzpicture}
        \begin{axis}[
            width=8cm,
            height=4cm,
            ybar,
            bar width=2.5pt,
            ylabel={Weight},
            yticklabel style={
              /pgf/number format/precision=3,
              /pgf/number format/fixed},
            xlabel={Layer},
            xtick=data,
            ytick={0, 0.05, 0.10, 0.15},
            ymin=0,
            ymax=0.17,
            legend columns = -1,
            legend style={at={(0.5,1.4)},anchor=north}
            ]
            \addplot [ybar, fill=blue, draw opacity=1] coordinates {  
                (1,0.08565809577703476) (2,0.07609087228775024) (3,0.06305480003356934) (4,0.060751307755708694) (5,0.0820808932185173) (6,0.09371702373027802) (7,0.11568929255008698) (8,0.12552812695503235) (9,0.09709346294403076) (10,0.0721869096159935) (11,0.06093427166342735) (12,0.0672149509191513)
            };
            \addplot [ybar, fill=red, draw opacity=1] coordinates {  
                (1,0.06983410567045212) (2,0.07103994488716125) (3,0.06852240115404129) (4,0.06579706817865372) (5,0.08125966042280197) (6,0.09709449112415314) (7,0.11736177653074265) (8,0.11956623196601868) (9,0.09743425250053406) (10,0.07187807559967041) (11,0.06529345363378525) (12,0.07491864264011383)
            };
            \addplot [ybar, fill=yellow, draw opacity=1] coordinates {  
                (1,0.05814964696764946) (2,0.06132270768284798) (3,0.05158567801117897) (4,0.043610941618680954) (5,0.06644365936517715) (6,0.08924068510532379) (7,0.1601545214653015) (8,0.1550816148519516) (9,0.11505492031574249) (10,0.06671051681041718) (11,0.0544387623667717) (12,0.07820641249418259)
            };
            \legend{Split 1,Split 2,Split 3}
        \end{axis}
    \end{tikzpicture}
    \caption{Learned weights for the primary model Whisper decoder layers.}
    \label{fig:whisper_layers}
\end{figure}

The learned weights for the Whisper decoder layers from the primary model are shown in \autoref{fig:whisper_layers}. These show how the model used the information to weigh each decoder layer feature, and therefore the higher the value the more useful the model finds the layer to be.  The weights are shown for each of the three models trained on the different training splits.  

The general pattern for the different training splits is similar, with layers $7$ and $8$ having the highest weights across all splits. This suggests that layers $7$ and $8$ contain the most relevant information for intelligibility. Interestingly, the model trained on Split 3 learns weights that emphasise layers 7 and 8 more strongly.
\subsection{Performance comparison}
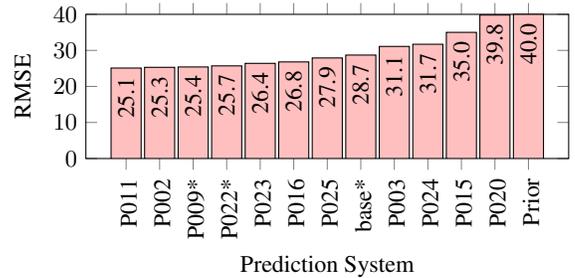
\begin{figure}[!ht]
    \centering
    \begin{tikzpicture}
        \begin{axis}[
            width=8cm,
            height=3.5cm,
            ymin=0,
            ymax=40,
            ybar,
            bar width=0.4cm,
            symbolic x coords={P011,P002,P009*,P022*,P023,P016,P025,base*,P003,P024,P015,P020,Prior},
            xtick={P011,P002,P009*,P022*,P023,P016,P025,base*,P003,P024,P015,P020,Prior},
            xticklabel style={text height=2ex,rotate=90},
            ylabel style={align=center, text height=8mm},
            yticklabel style={text width=4mm, align=right},
            ylabel={RMSE},
            xlabel={Prediction System},
            point meta=explicit symbolic, 
            visualization depends on=\thisrow{alignment} \as \alignment,
            nodes near coords,
            every node near coord/.append style={rotate=90, anchor=east}
]
            \addplot[ybar, fill=pink]
            table[meta index=2] {
                x y label alignment
                P011 25.1 25.1 180
                P002 25.3 25.3 180
                P009* 25.4 25.4 180
                P022* 25.7 25.7 180
                P023 26.4 26.4 180
                P016 26.8 26.8 180
                P025 27.9 27.9 180
                base* 28.7 28.7 180
                P003 31.1 31.1 180
                P024 31.7 31.7 180
                P015 35.0 35.0 0
                P020 39.8 39.8 0
                Prior 40.0 40.0 0
            };

        \end{axis}
    \end{tikzpicture}
    \caption{Comparison of performance of CPC2 entries. A * denotes an intrusive system.}
    \label{fig:challenge-performance}
\end{figure}

\autoref{fig:challenge-performance} shows the performance of the proposed model (\emph{P002}) compared to all other challenge entries, as well as the challenge \ac{HASPI} baseline. \emph{Prior} is a system which always predicts the average of the intelligibility over the training set, regardless of the input. The proposed system is outperformed by only one other system, P011, which also utilises Whisper-derived features. The difference in performance is very slight, with P011 achieving an RMSE score of $25.1$, while our system achieves $25.3$.

\section{Conclusions}
\label{sec:Conclusions}
We show that Whisper decoder layers are a useful feature representation for speech intelligibility prediction, with layers $7$ and $8$ appearing to be the most relevant.

Our proposed system performs substantially better than the \ac{HASPI} regression baseline and all but one of the other challenge approaches, even outperforming intrusive systems which had access to the clean reference signal. It is able to generalise to unseen enhancement systems and listeners.


\vfill\pagebreak

\bibliographystyle{IEEEbib-abbrev}
\bibliography{strings,refs}

\begin{thebibliography}{10}

\bibitem{world2018addressing}
{World Health Organization},
\newblock ``{Addressing the Rising Prevalence of Hearing Loss},''
\newblock 2018,
\newblock ISBN: 9789241550260.

\bibitem{RGA11}
J.~Rennies, S.~Goetze, and J.-E. Appell,
\newblock ``{P}ersonalized {A}coustic {I}nterfaces for {H}uman-{C}omputer
  {I}nteraction,''
\newblock in {\em {Human-Centered Design of E-Health Technologies: Concepts,
  Methods and Applications}}, M.~Ziefle and C.R{\"{o}}cker, Eds., chapter~8,
  pp. 180--207. {IGI} {G}lobal, 2011.

\bibitem{aging_stats}
{World Health Organisation},
\newblock ``{Ageing and Health},''
  \url{https://www.who.int/news-room/fact-sheets/detail/ageing-and-health},
\newblock Accesssed: 2023-07-26.

\bibitem{Doclo_HA_Overview}
S.~Doclo, W.~Kellermann, S.~Makino, and S.~E. Nordholm,
\newblock ``{Multichannel Signal Enhancement Algorithms for Assisted Listening
  Devices: Exploiting spatial diversity using multiple microphones},''
\newblock {\em IEEE Signal Processing Magazine}, vol. 32, no. 2, pp. 18--30,
  2015.

\bibitem{GXRRA10}
S.~Goetze, F.~Xiong, J.~Rennies, T.~Rohdenburg, and J.~Appell,
\newblock ``{Hands-Free Telecommunication for Elderly Persons Suffering from
  Hearing Deficiencies},''
\newblock in {\em IEEE Int.~Conf.~on E-Health Networking, Application and
  Services (Healthcom'10)}, 2010.

\bibitem{falk2015objective}
T.~H. Falk, V.~Parsa, J.~F. Santos, K.~Arehart, O.~Hazrati, R.~Huber, J.~M.
  Kates, and S.~Scollie,
\newblock ``{Objective Quality and Intelligibility Prediction for Users of
  Assistive Listening Devices: Advantages and Limitations of Existing Tools},''
\newblock {\em IEEE Signal Processing Magazine}, vol. 32, no. 2, pp. 114--124,
  2015.

\bibitem{WKJ+14}
A.~Warzybok, I.~Kodrasi, J.~Jungmann, E.~Habets, T.~Gerkmann, A.~Mertins,
  S.~Doclo, B.~Kollmeier, and S.~Goetze,
\newblock ``{Subjective Speech Quality and Speech Intelligibility Evaluation of
  Single-Channel Dereverberation Algorithms},''
\newblock in {\em Int. Workshop on Acoustic Signal Enhancement (IWAENC 2014)},
  France, Sep. 2014.

\bibitem{Voelker_SI_2015}
C.~Völker, A.~Warzybok, and S.~Ernst,
\newblock ``{C}omparing {B}inaural {P}re-processing {S}trategies {III}:
  {S}peech {I}ntelligibility of {N}ormal-{H}earing and {H}earing-{I}mpaired
  {L}isteners,''
\newblock {\em Trends in Hearing}, vol. 19, 2015.

\bibitem{CSS+19}
B.~{Cauchi}, K.~{Siedenburg}, J.~F. {Santos}, T.~H. {Falk}, S.~{Doclo}, and
  S.~{Goetze},
\newblock ``{N}on-{I}ntrusive {S}peech {Q}uality {P}rediction {U}sing
  {M}odulation {E}nergies and {LSTM}-{N}etwork,''
\newblock {\em IEEE/ACM Transactions on Audio, Speech, and Language
  Processing}, vol. 27, no. 7, pp. 1151--1163, July 2019.

\bibitem{barker24_icassp}
J.~Barker, M.~Akeroyd, W.~Bailey, T.~J. Cox, J.~F. Culling, J.~Firth,
  S.~Graetzer, and G.~Naylor,
\newblock ``{The 2nd Clarity Prediction Challenge: A machine learning challenge
  for hearing aid intelligibility prediction},''
\newblock in {\em ICASSP}, 2024.

\bibitem{barker22_interspeech}
J.~Barker, M.~Akeroyd, T.~J. Cox, J.~F. Culling, J.~Firth, S.~Graetzer,
  H.~Griffiths, L.~Harris, G.~Naylor, Z.~Podwinska, E.~Porter, and R.~V. Munoz,
\newblock ``{The 1st Clarity Prediction Challenge: A machine learning challenge
  for hearing aid intelligibility prediction},''
\newblock in {\em Proc. Interspeech}, 2022, pp. 3508--3512.

\bibitem{tu22b_interspeech}
Z.~Tu, N.~Ma, and J.~Barker,
\newblock ``{Unsupervised Uncertainty Measures of Automatic Speech Recognition
  for Non-intrusive Speech Intelligibility Prediction},''
\newblock in {\em Proc. Interspeech}, 2022, pp. 3493--3497.

\bibitem{zezario2022mbinet}
R.~E. Zezario, F.~Chen, C.-S. Fuh, H.-M. Wang, and Y.~Tsao,
\newblock ``{MBI-Net: A Non-Intrusive Multi-Branched Speech Intelligibility
  Prediction Model for Hearing Aids},'' 2022.

\bibitem{close2023non}
G.~Close, T.~Hain, and S.~Goetze,
\newblock ``{Non Intrusive Intelligibility Predictor for Hearing Impaired
  Individuals using Self Supervised Speech Representations},''
\newblock in {\em Proc.\ Workshop on Speech Foundation Models and their
  Performance Benchmarks (SPARKS), ASRU sattelite workshop}, Taipei, Taiwan,
  2023.

\bibitem{close22_interspeech}
G.~Close, S.~Hollands, T.~Hain, and S.~Goetze,
\newblock ``{Non-intrusive Speech Intelligibility Metric Prediction for Hearing
  Impaired Individuals},''
\newblock in {\em Proc. Interspeech}, 2022, pp. 3483--3487.

\bibitem{kates2021hearing}
J.~M. Kates and K.~H. Arehart,
\newblock ``{The Hearing-aid Speech Perception Index (HASPI) Version 2},''
\newblock {\em Speech Communication}, vol. 131, pp. 35--46, 2021.

\bibitem{whisper}
A.~Radford, J.~W. Kim, T.~Xu, G.~Brockman, C.~McLeavey, and I.~Sutskever,
\newblock ``{Robust Speech Recognition via Large-Scale Weak Supervision},''
  2022.

\bibitem{vaswani2017attention}
A.~Vaswani, N.~Shazeer, N.~Parmar, J.~Uszkoreit, L.~Jones, A.~N. Gomez,
  {\L}.~Kaiser, and I.~Polosukhin,
\newblock ``{Attention is All You Need},''
\newblock {\em Advances in neural information processing systems}, vol. 30,
  2017.

\bibitem{nisqa_pretrained_ss}
B.~Tamm, H.~Balabin, R.~Vandenberghe, and H.~V. hamme,
\newblock ``{Pre-trained Speech Representations as Feature Extractors for
  Speech Quality Assessment in Online Conferencing Applications},''
\newblock in {\em Interspeech}. Sep 2022, {ISCA}.

\bibitem{hintzman1984minerva2}
D.~Hintzman,
\newblock ``{MINERVA 2: a Simulation Model of Human Memory},''
\newblock {\em Behaviour Research Methods, Instruments \& Computers}, vol. 16,
  pp. 96--101, 03 1984.

\bibitem{psyEx_rulePlusEx}
R.~M. Nosofsky, T.~J. Palmeri, and M.~Stephen~C,
\newblock ``{Rule-Plus-Exception Model of Classification Learning},''
\newblock {\em Psychological Review}, vol. 101, no. 1, pp. 53--79, 1994.

\bibitem{humanRuleExemp:1998}
M.~A. Erickson and J.~K. Kruschke,
\newblock ``{Rules and Exemplars in Category Learning},''
\newblock {\em Journal of Experimental Psychology: General}, vol. 127, 1998.

\bibitem{humanExProt:2006}
J.~N. Rouder and R.~Ratcliff,
\newblock ``{Comparing Exemplar- and Rule-Based Theories of Categorization},''
\newblock {\em Current Directions in Psychological Science}, vol. 15, 2006.

\bibitem{psyPersonaSwitch}
S.~Natal, I.~McLaren, and E.~Livesey,
\newblock ``{Generalization of Feature- and Rule-based Learning in the
  Categorization of Dimensional Stimuli: Evidence for Dual Processes Under
  Cognitive Control},''
\newblock {\em J Exp Psychol Anim Behav Process}, vol. 39, no. 2, pp. 140--51,
  2013.

\bibitem{better_ear}
I.~Gibbs, Bobby~E., J.~G.~W. Bernstein, D.~S. Brungart, and M.~J. Goupell,
\newblock ``{Effects of better-ear glimpsing, binaural unmasking, and spectral
  resolution on spatial release from masking in cochlear-implant users},''
\newblock {\em The Journal of the Acoustical Society of America}, vol. 152, no.
  2, pp. 1230--1246, 08 2022.

\end{thebibliography}

\end{document}